\documentclass[sn-mathphys-num]{sn-jnl}% Math and Physical Sciences Numbered Reference Style 
\usepackage{amsmath}
\usepackage{scalerel}
\usepackage{stackengine,wasysym}

\newcommand\reallywidetilde[1]{\ThisStyle{%
  \setbox0=\hbox{$\SavedStyle#1$}%
  \stackengine{-.1\LMpt}{$\SavedStyle#1$}{%
    \stretchto{\scaleto{\SavedStyle\mkern.2mu\AC}{.5150\wd0}}{.6\ht0}%
  }{O}{c}{F}{T}{S}%
}}

\usepackage{graphicx}%
\usepackage{multirow}%
\usepackage{amsfonts}%

\usepackage{mathrsfs}%
\usepackage[title]{appendix}%
\usepackage{xcolor}%
\usepackage{textcomp}%
\usepackage{manyfoot}%
\usepackage{booktabs}%
\usepackage{algorithm}%
\usepackage{algorithmicx}%
\usepackage{algpseudocode}%
\usepackage{listings}%
\usepackage[T1]{fontenc}
\usepackage{amssymb}
\usepackage{mathtools}

\usepackage{bm}
\usepackage{stackengine}
\usepackage{scalerel}
\usepackage{framed}
\usepackage{mathtools}

\usepackage{color}
\usepackage{bbold}
\usepackage{bm}
\usepackage{latexsym}
\usepackage{braket}
\usepackage{slashed}
\begin{document}

\title{\boldmath Massless limit and conformal soft limit for celestial massive amplitudes\unboldmath}

\author{\fnm{Wei} \sur{Fan}}
\affil{\orgdiv{Department of Physics, School of Science}, \orgname{Jiangsu University of Science and Technology}, \orgaddress{ \city{Zhenjiang}, \postcode{212114},  \country{China}}}

% E-mail addresses: only for the corresponding author
\email{fanwei@just.edu.cn}

\abstract{
In celestial holography, the  massive and massless scalars in 4d space-time are represented by the Fourier transform of the bulk-to-boundary propagators and the Mellin transform of plane waves respectively. Recently, the 3pt celestial amplitude of one massive scalar and two massless scalars was discussed in arXiv:2312.08597. In this paper, we compute the 3pt celestial amplitude of two massive scalars and one massless scalar. Then we take the massless limit $m\to 0$ for one of the massive scalars, during which process the gamma function $\Gamma(1-\Delta)$ appears. By requiring the resulting amplitude to be well-defined, that is it goes to the 3pt amplitude of arXiv:2312.08597, the scaling dimension  of this massive scalar has to be conformally soft $\Delta \to 1$. The pole $1/(1-\Delta)$ coming from $\Gamma(1-\Delta)$ is crucial for this massless limit. Without it the resulting amplitude would be zero.  This can be compared with the conformal soft limit in celestial gluon amplitudes, where a singularity $1/(\Delta -1)$ arises and the leading contribution comes from the soft energy $\omega\to 0$. The phase factors in the massless limit of massive conformal primary wave functions, discussed in arXiv:1705.01027, plays an import and consistent role in the celestial massive amplitudes. Furthermore, the subleading orders $m^{2n}$ can also contribute poles when the scaling dimension is  analytically continued to $\Delta=1-n$ or $\Delta = 2$, and we find that this consistent massless limit only exists for dimensions belonging to the generalized conformal primary operators $\Delta \in 2-\mathbb{Z}_{\geqslant 0}$ of massless bosons.
}


\maketitle

\tableofcontents

\section{Introduction}
\label{sec:intro}

Celestial conformal field theory (CCFT)  is a flat holography~\cite{Strominger:2017zoo,Pasterski:2021raf, Pasterski:2021rjz,Raclariu:2021zjz} of Minkowski spacetime.  The bulk QFT in Minkowski spacetime $\mathbb{R}^{1,3}$ can be recast as a boundary CFT in the celestial sphere. The amplitudes are converted from the  momentum basis to the  boost basis (the so-called conformal primary wavefunction)~\cite{Pasterski:2017kqt} and then become the CFT correlation functions. The consequence of 4d soft theorem to celestial holography is one of the import problems. For example, the soft energy $\omega\to 0$ generates the leading contribution to the conformal soft limit $\Delta\to 1$ in  celestial gluon amplitudes~\cite{Cheung:2016iub,Fan:2019emx,Nandan:2019jas,Pate:2019mfs}. 

A natural question to ask is what would be the corresponding physics related with the conformal soft limit of celestial massive amplitudes? We address this question in this paper using the 3pt celestial amplitude of two massive scalars and one massless scalar. We find that the massless limit $m\to 0$ of this amplitude leads to the conformal soft limit $\Delta\to 1$, to make the resulting amplitude well-defined, see~\eqref{eq:core}. On the basis level, the massless limit of the conformal primary wave functions has a phase factor $m^{-\Delta}$~\cite{Pasterski:2017kqt}. After going to the celestial amplitudes, the massless limit is accompanied by the gamma function $\Gamma(1-\Delta)$, which contributes a pole in the conformal soft limit $\Delta\to 1$. This resembles, in spirit, to the conformal soft limit of celestial gluons, which appears in their amplitudes rather than  conformal primary waves.

Note that in this paper the 'conformally soft' is associated with the massless scalar after taking the massless limit. For the massive scalar, $\Delta \to 1$ is only choosing a specific dimension, but after the massless limit $m\to 0$ it becomes the conformal soft dimension. Because the specific choice of dimension and the massless limit are taken together in this paper, we will directly use 'conformally soft' for simplicity.

The celestial massive amplitudes are hard to compute in 4d space-time. In the original paper of massive conformal primary waves~\cite{Pasterski:2016qvg}, the 3pt celestial amplitudes of three massive scalars was computed in the near-extremal limit $m_1=2(1+\epsilon)m,m_2=m_3=m$. The result is the tree-level 3pt Witten diagram in the leading order of $\sqrt{\epsilon}$ when $\epsilon\to 0$. The 3pt celestial amplitude of one massive scalar and two massless scalars  $\langle \Phi_{\Delta_1}^{m_1} \phi_{\Delta_2} \phi_{\Delta_3}\rangle$ was computed recently by~\cite{Himwich:2023njb}\footnote{This kind of celestial amplitudes was firstly discussed by ~\cite{Lam:2017ofc} in  3d space-time, for the implications of the 4d optical theorem on celestial holography.}, where the non-local behavior of massive states on celestial sphere was discussed. Since massive states are highly nontrivial in celestial holography, it is meaningful to investigate the physics related with their conformal soft limit. Here we continue to compute the 3pt celestial amplitude of two massive scalars and one massless scalar  $\langle \Phi_{\Delta_1}^{m_1} \Phi_{\Delta_2}^{m_2} \phi_{\Delta_3}\rangle$. Using the standard method of hyperbolic coordinates and  Feynman parametrization, the two masses $m_1$ and $m_2$ appear only in the finite integral of the Feynman parameters. Then we can safely take the massless limit $m_2\to 0$. Combined with the conformal soft limit $\Delta_2\to 1$, the 3pt massive amplitude $\langle \Phi_{\Delta_1}^{m_1} \Phi_{\Delta_2}^{m_2} \phi_{\Delta_3}\rangle$ has the desired limit $\langle \Phi_{\Delta_1}^{m_1} \phi_{\Delta_2} \phi_{\Delta_3}\rangle$. The general result of this amplitude is given in~\eqref{eq:totAmp}.

On the basis level, the massless limit of the massive conformal primary wave function leads to two massless fields $\Phi_{\Delta_2}^{m}\to \phi_{\Delta_2}, \reallywidetilde{\phi_{\tilde{\Delta}_2}}$~\cite{Pasterski:2017kqt}, which are the same for $\Delta_2=1$. They are different for general values of $\Delta_2$. Using the general result of this amplitude  in~\eqref{eq:totAmp}, we can study subleading orders $m_2^{2n}$ of the massless limit. Then we find that it leads to the 3pt amplitude $\langle \Phi_{\Delta_1}^{m_1} \phi_{\Delta_2} \phi_{\Delta_3}\rangle$ for the analytically continued dimension $\Delta_2 = 1-n$, see~\eqref{eq:subordermatch}, and it leads to the 3pt amplitude $\langle \Phi_{\Delta_1}^{m_1} \reallywidetilde{\phi_{\tilde{\Delta}_2}} \phi_{\Delta_3}\rangle$ for the analytically continued dimension $\Delta_2 =2$, see~\eqref{eq:shadowlimit}. For analytically continued dimension $\Delta_2 \ge 3$, there is a massless limit but it is neither $\langle \Phi_{\Delta_1}^{m_1} \reallywidetilde{\phi_{\tilde{\Delta}_2}} \phi_{\Delta_3}\rangle$ nor $\langle \Phi_{\Delta_1}^{m_1} \phi_{\Delta_2} \phi_{\Delta_3}\rangle$. So it is only for dimensions $\Delta_2 \le 2$ that the massless limit of this amplitude is consistent. These dimensions happen to belong to the range of generalized conformal primary operators $\Delta \in 2-\mathbb{Z}_{\geqslant 0}$ of massless bosons~\cite{Mitra:2024ugt}. So the consistent massless limit of $\langle \Phi_{\Delta_1}^{m_1} \Phi_{\Delta_2}^{m_2} \phi_{\Delta_3}\rangle$ picks up these generalized conformal primary operators.
Note that this amplitude was also discussed recently in~\cite{Liu:2024lbs},  whose results and physics content have no overlap with us. 

Without going into the detailed algebra of ~\eqref{eq:totAmp}, there is a simple explanation of this 'asymmetry' between the two massless  fields $\phi_{\Delta_2}, \reallywidetilde{\phi_{\tilde{\Delta}_2}}$ in the massless limit of $\Phi_{\Delta_2}^{m}$. Inside the amplitude of $\Phi_{\Delta_2}^{m}$, some  poles of $\Delta_2$ can be rewritten in terms of poles of $\tilde{\Delta}_2$. But the two dimensions $\Delta_2, \tilde{\Delta}_2$ are not equal, except at the value of $1$. So $\Delta_2$ and $\tilde{\Delta}_2$ can not appear in a 'symmetric' way inside the amplitude of $\Phi_{\Delta_2}^{m}$,  whatever the 'rewritting' is. This point can be understood via proof by contradition. If  $\Delta_2$ and $\tilde{\Delta}_2$ appear symmetrically inside amplitudes of $\Phi_{\Delta_2}^{m}$, then it means amplitudes of $\Phi_{\Delta_2}^{m}$ equals amplitudes of $\Phi_{\tilde{\Delta}_2}^{m}$. But  amplitudes of $\Phi_{\Delta_2}^{m}$ and  amplitudes of $\Phi_{\tilde{\Delta}_2}^{m}$ do not equal generally, because $\phi_{\Delta_2}$ and $\reallywidetilde{\phi_{\tilde{\Delta}_2}}$ are not equal. 

In a word, it is because we are starting from the amplitude of $\Phi_{\Delta_2}^{m}$ that only $\Delta_2 \in 2-\mathbb{Z}_{\geqslant 0}$ can give a consistent massless limit. If we start from the amplitude of $\Phi_{\tilde{\Delta}_2}^{m}$, we would get $\tilde{\Delta}_2 \in 2-\mathbb{Z}_{\geqslant 0}$ as the consistent massless limit. This is somehow tautological. 

The paper is organized as follows. In Section 2, we give a short account of the notation and briefly review the 3pt celestial amplitude  $\langle \Phi_{\Delta_1}^{m_1} \phi_{\Delta_2} \phi_{\Delta_3}\rangle$. Then we compute the shadowed 3pt amplitude $\langle \Phi_{\Delta_1}^{m_1} \reallywidetilde{\phi_{\tilde{\Delta}_2}} \phi_{\Delta_3}\rangle$.  In Section 3, we compute the 3pt celestial amplitude  $\langle \Phi_{\Delta_1}^{m_1} \Phi_{\Delta_2}^{m_2} \phi_{\Delta_3}\rangle$ using the Feynman parametrization. The complete result of this amplitude is put to the appendix.  In Section 4, we investigate the massless limit $m_2\to 0$ and the conformal soft limit $\Delta_2\to 1$. Then we discuss the subleading terms $m_2^{2n}$ and the analytic continuation to $\Delta_2=1-n$ and $\Delta_2=2$.   We conclude with a discussion of open questions  in Section 5.

\section{Preliminaries}

In CCFT, the massless four-momentum $p_i^\mu=\epsilon_i \omega_i \hat{q}_i^\mu$ is parameterized by the energy $\omega_i$ and the point $(w_i,\bar{w}_i)$ of the celestial sphere via the formula~\cite{Pasterski:2017kqt} 
\begin{align}
  \hat{q}_i =   (1+w_i\bar{w}_i, w_i+\bar{w}_i,  -i(w_i-\bar{w}_i),1-w_i\bar{w}_i),
\end{align}
 where  $\epsilon_i=\pm 1$ represents outgoing/incoming momentum. During the computation, we use the coordinate $\vec{w}_i=(\operatorname{Re}(w_i),\operatorname{Im}(w_i))$ for the celestial sphere, in order to save space. Each 4d massless scalar corresponds to a scalar conformal primary wave function $\phi_{\Delta}(X^\mu;\vec{w})$ (the so-called boost basis) via the Mellin transform 
\begin{align}
  \label{eq:mellin}
  \phi_{\Delta}^{(\pm)}(X^\mu;\vec{w})\coloneqq \int_0^\infty d\omega \omega^{\Delta-1} e^{\pm i  \omega \hat{q}(\vec{w})\cdot X -\varepsilon \omega},
\end{align}
where $\varepsilon>0$ is a regularization parameter and the conformal dimensions are $h_i=\bar{h}_i=\Delta_i/2=(1 + i\lambda_i)/2, \lambda_i\in \mathbb{R}$. 

The massive four-momentum $p_i^\mu=\epsilon_i m_i \hat{p}^\mu$ is parameterized by the mass of the particle $m_i$ and the hyperbolic coordinate $(y_i, \vec{z}_i)$ via the formula~\cite{Pasterski:2016qvg}
\begin{align}
  \hat{p}_i^\mu(y_i, \vec{z}_i)=\left(\frac{1+y_i^2+|z_i|^2}{2 y_i}, \frac{\vec{z}_i}{y_i},  \frac{1-y_i^2-|z_i|^2}{2 y_i}\right).
\end{align}
The massive scalar corresponds to a scalar conformal primary wave $\Phi_{\Delta}^{m}(X^\mu;\vec{w})$   via the Fourier transform 
\begin{align}
  \Phi_{\Delta}^{m(\pm)}(X^\mu;\vec{w})\coloneqq \int_0^\infty \frac{dy}{y^3}\int d^2z G_{\Delta}(y,\vec{z};\vec{w})e^{\pm i m \hat{p}(y,\vec{z})\cdot X},
\end{align}
where $G_{\Delta}(y,\vec{z};\vec{w})$ is the bulk-to-boundary propagator with scaling dimension $\Delta$ 
\begin{align}
G_{\Delta}(\hat{p}(y,\vec{z}) ; \hat{q}(\vec{w}))=\frac{1}{(-\hat{p} \cdot \hat{q})^{\Delta}}=\left(\frac{y}{y^2+|\vec{z}-\vec{w}|^2}\right)^{\Delta}.
\end{align}

The Lorentz  transformation $\Lambda^\mu_\nu \in S O(1, 3)$ acts non-linearly on the coordinates $\vec{w} \to \vec{w}^{\prime}(\vec{w}), \vec{z} \to \vec{z}^{\prime}(y, \vec{z}), y \to y^{\prime}(y, \vec{z})$, but the four-momenta transform linearly
\begin{align}
  \hat{p}^\mu\left(y^{\prime}, \vec{z}^{\prime}\right)=\Lambda^\mu{ }_\nu \hat{p}^\nu, \, q^\mu\left(\vec{w}^{\prime}\right)=\left|\frac{\partial \vec{w}^{\prime}}{\partial \vec{w}}\right|^{1 / 2} \Lambda_\nu^\mu q^\nu(\vec{w}).
\end{align}
Then the conformal primary waves have the conformal symmetry on the celestial sphere
\begin{align}
\label{eq:sl2c}
\Big\{\Phi_{\Delta}^{m},\phi_{\Delta}\Big\}(\Lambda_\nu^\mu X^\nu ; \vec{w}^{\prime}(\vec{w}))=\left|\frac{\partial \vec{w}^{\prime}}{\partial \vec{w}}\right|^{-\Delta / 2} \Big\{\Phi_{\Delta}^{m},\phi_{\Delta}\Big\}(X^\mu ; \vec{w}),
\end{align}
where we write $\Phi_{\Delta}^{m}(X  ; \vec{w})$ and $\phi_{\Delta}(X  ; \vec{w})$ together to save space.

In the massless limit, the massive conformal primary wavefunction $\Phi_{\Delta}^{m(\pm)}(X^\mu;\vec{w})$ has the following behavior~\cite{Pasterski:2017kqt}
\begin{align}
  \label{eq:basislimit}
  \Phi_{\Delta}^{m}(X ; \vec{w}) \overset{m \rightarrow 0}{\longrightarrow}\left(\frac{m}{2}\right)^{-\Delta} \frac{\pi\Gamma(\Delta-1)}{\Gamma\left(\Delta\right)} \phi_{\Delta}(X^\mu;\vec{w}) + \left(\frac{m}{2}\right)^{-\tilde{\Delta}} \frac{\pi\Gamma(\tilde{\Delta}-1)}{\Gamma\left(\tilde{\Delta}\right)}\reallywidetilde{\phi_{\tilde{\Delta}}}(X^\mu;\vec{w}),
\end{align}
where $\tilde{\Delta}=2-\Delta$ is the shadow dimension and $\reallywidetilde{\phi_{\tilde{\Delta}}}$ is the shadow operator of $\phi_{\tilde{\Delta}}$. In the basis level, the massless limit is not well-defined, due to this phase factor $m^{-\Delta}$. In the amplitude level, however, things are different. The physical interaction between particles makes the massless limit well-defined, provided that the conformal soft limit $\Delta \to 1$ is together with the massless limit $m\to 0$. This is what we will show in this paper. It turns out that the phase factor $m^{-\Delta}$ is crucial for the massless limit of celestial amplitudes.

\subsection{The 3pt celestial amplitude $\langle \Phi_{\Delta_1}^{m_1} \phi_{\Delta_2} \phi_{\Delta_3}\rangle$ and $\langle \Phi_{\Delta_1}^{m_1} \widetilde{\phi_{\tilde{\Delta}_2}} \phi_{\Delta_3}\rangle$}

Here we briefly review the 3pt celestial amplitude  $\langle \Phi_{\Delta_1}^{m_1} \phi_{\Delta_2} \phi_{\Delta_3}\rangle$ appearing in~\cite{Himwich:2023njb}. This 3pt  amplitude is given by the following integral
\begin{align}
  \langle \Phi_{\Delta_1}^{m_1}(\vec{w}_1) \phi_{\Delta_2}(\vec{w}_2) \phi_{\Delta_3}(\vec{w}_3)\rangle=i \lambda \int d^4 X\, \Phi_{\Delta_1}^{m_1(-)}\left(X^\mu ; \vec{w}_1\right) \prod_{i=2}^3 \phi_{\Delta_i}^{(+)}\left(X^\mu ; \vec{w}_i \right),
\end{align}
where $\Phi_{\Delta_i}^{m_i}(\vec{w}_i)$ are the corresponding conformal primary operators in CCFT. The integration over $X$ can be computed first, which leads to a momentum conservating delta function
\begin{align}
  \int \frac{d y_1}{y_1^3} \int d^2 z_1 \left(\frac{y_1}{y_1^2+|\vec{z}_1-\vec{w}_1|^2}\right)^{\Delta_1} \int \left(\prod_{i=2}^{3}\frac{d \omega_i}{\omega_i} \omega_i^{\Delta_i} \right)\delta^4(-m_1 \hat{p}_1+ \omega_2\hat{q}_2+\omega_3\hat{q}_3).
\end{align}
This momentum conservating delta function can be used to fix the integration variables $(y_1, \vec{z}_1, \omega_2)$, then only one integration $\omega_3$ is left, which can be done analytically. The final result is
\begin{align}
  \label{eq:Himwich}
  \frac{i\lambda (2\pi)^4 }{16}\left(\frac{m_1}{2}\right)^{\Delta_2+\Delta_3-4}  \frac{\Gamma(\frac{\Delta_1+\Delta_2-\Delta_3}{2}) \Gamma(\frac{\Delta_1-\Delta_2+\Delta_3}{2})/\Gamma(\Delta_1)}{|\vec{w}_{12}|^{\Delta_1+\Delta_2-\Delta_3}|\vec{w}_{13}|^{\Delta_1+\Delta_3-\Delta_2}|\vec{w}_{23}|^{\Delta_2+\Delta_3-\Delta_1}}, 
\end{align}
where $\vec{w}_{ij}\coloneqq \vec{w}_i - \vec{w}_j$ and the overall constant are adapted to the notation of this paper. 

In the following, we continue to compute the 3pt amplitude with two massive scalars and one massless scalar  $\langle \Phi_{\Delta_1}^{m_1} \Phi_{\Delta_2}^{m_2} \phi_{\Delta_3}\rangle$. By taking  one  massive scalar to be massless $m_2\to 0$, we show that it reduces to the 3pt amplitude $\langle \Phi_{\Delta_1}^{m_1} \phi_{\Delta_2} \phi_{\Delta_3}\rangle$ as expected. The massless limit of $\Phi_{\Delta_2}^{m}(X ; \vec{w})$ contains both $\phi_{\Delta_2}$ and $\reallywidetilde{\phi_{\tilde{\Delta}_2}}$, so we also need the 3pt amplitude $\langle \Phi_{\Delta_1}^{m_1}\reallywidetilde{\phi_{\tilde{\Delta}_2}} \phi_{\Delta_3}\rangle$. To do that, we need the shadow transformation. 

The shadow of a primary operator $\phi_{\Delta}(z, \bar{z})$ with conformal dimension $h,\bar{h}$, hence with the scaling dimension $\Delta=h+\bar{h}$ and spin
$J=h-\bar{h}$, is defined \cite{Osborn:2012vt} as
\begin{align}
\label{eq:shadowDef}
\tilde{\phi}_{\tilde{\Delta}}(z, \bar{z})\coloneqq\reallywidetilde{\phi_{\Delta}(y, \bar{y})}=k_{h, \bar{h}} \int d^2 y(z-y)^{2 h-2}(\bar{z}-\bar{y})^{2 \bar{h}-2} \phi_{\Delta}(y, \bar{y}),    
\end{align}
where the constant $ k_{h, \bar{h}}=(-1)^{2(h-\bar{h})}\Gamma(2-2 h)/(\pi \Gamma(2 \bar{h}-1)) $ is chosen in such a way that, for integer or half-integer spin, $\tilde{\tilde\phi}(z,\bar{z})=\phi(z,\bar{z})$. 
The shadow field $\tilde{\phi}_{\tilde{\Delta}}(z,\bar{z})$ is  a primary operator with conformal dimension $1-h,1-\bar{h}$, hence with $\tilde{\Delta}=2-\Delta$ and  $ \tilde J=-J$. 

Now we can perform the shadow transform on $\phi_{\tilde{\Delta}_2}$ with conformal dimension $h=\bar{h}=\tilde{\Delta}_2/2$ to get the shadowed field $\reallywidetilde{\phi_{\tilde{\Delta}_2}}(\vec{w}_2) = \reallywidetilde{\phi_{\tilde{\Delta}_2}(\vec{w}'_2)}$  
\begin{align}
\label{eq:shadowAmp}
& \langle \Phi_{\Delta_1}^{m_1}(\vec{w}_1) \reallywidetilde{\phi_{\tilde{\Delta}_2}}(\vec{w}_2)  \phi_{\Delta_3}(\vec{w}_3)\rangle=  k_{h, \bar{h}}\int \frac{d^2\vec{w}'_2}{|\vec{w}'_2-\vec{w}_2|^{4-2\tilde{\Delta}_2}} \langle \Phi_{\Delta_1}^{m_1}(\vec{w}_1) \phi_{\tilde{\Delta}_2}(\vec{w}'_2) \phi_{\Delta_3}(\vec{w}_3)\rangle.
\end{align}
Using the conformal symmetry, we shall perform a conformal transformation to set $\vec{w}_1\to 0, \vec{w}_2\to  1, \vec{w}_3\to \infty $,  $\vec{w}'_2\to \vec{x}$ and obtain the function
\begin{align}
  \langle \Phi_{\Delta_1}^{m_1}(\vec{w}_1) & \reallywidetilde{\phi_{\tilde{\Delta}_2}}(\vec{w}_2)  \phi_{\Delta_3}(\vec{w}_3)\rangle =   \frac{i\lambda (2\pi)^4 m_1^{\tilde{\Delta}_2+\Delta_3-4}}{2^{\tilde{\Delta}_2+\Delta_3}  |\vec{w}_{12}|^{\Delta_1+\Delta_2-\Delta_3}|\vec{w}_{13}|^{\Delta_1+\Delta_3-\Delta_2}|\vec{w}_{23}|^{\Delta_2+\Delta_3-\Delta_1} } \nonumber\\
  & \times \frac{\Gamma(2-\tilde{\Delta}_2)\Gamma(\frac{\Delta_1+\tilde{\Delta}_2-\Delta_3}{2}) \Gamma(\frac{\Delta_1-\tilde{\Delta}_2+\Delta_3}{2})}{\pi\Gamma(\tilde{\Delta}_2-1)\Gamma(\Delta_1)} \int\, d^2\vec{x}\, |\vec{x}|^{\Delta_3 -\Delta_1 -\tilde{\Delta}_2} |\vec{x}-1|^{2\tilde{\Delta}_2-4 } .
\end{align}

Such kind of 2d scalar integrals can be evaluated by analytic continuation of $x$ on the complex plane
\begin{align}
\int\, d^2\vec{x}\, |\vec{x}|^{2 a } |\vec{x}-1|^{2b}=\pi \frac{\Gamma(1+a) \Gamma(1+b) \Gamma(-1-a-b)}{\Gamma(-a) \Gamma(-b) \Gamma(2+a+b)},
\end{align}
the details of which can be seen in Chapter 7 of~\cite{dotsenko:notes}. 

So we finally obtain the 3pt amplitude
\begin{align}
  \label{eq:3ptshadow}
  & \langle \Phi_{\Delta_1}^{m_1}(\vec{w}_1) \reallywidetilde{\phi_{\tilde{\Delta}_2}}(\vec{w}_2)  \phi_{\Delta_3}(\vec{w}_3)\rangle =   \frac{i\lambda (2\pi)^4 m_1^{\tilde{\Delta}_2+\Delta_3-4}}{2^{\tilde{\Delta}_2+\Delta_3}}  \frac{ \frac{\Gamma(\frac{\Delta_1+\Delta_2-\Delta_3}{2}) \Gamma(\frac{\Delta_2+\Delta_3-\Delta_1}{2})}{\Gamma(\Delta_1)} \frac{\Gamma(-1+\frac{\Delta_1+\Delta_2+\Delta_3}{2})}{\Gamma(1+\frac{\Delta_3-\Delta_1-\Delta_2}{2})}}{ |\vec{w}_{12}|^{\Delta_1+\Delta_2-\Delta_3}|\vec{w}_{13}|^{\Delta_1+\Delta_3-\Delta_2}|\vec{w}_{23}|^{\Delta_2+\Delta_3-\Delta_1} }.
\end{align}

\section{The 3pt celestial amplitude $\langle \Phi_{\Delta_1}^{m_1} \Phi_{\Delta_2}^{m_2} \phi_{\Delta_3}\rangle$}

For convenience we choose the 1st particle to be incoming, the 2nd and the 3rd to be outgoing. The 3pt celestial amplitude  $\langle \Phi_{\Delta_1}^{m_1}  \Phi_{\Delta_2}^{m_2}  \phi_{\Delta_3}\rangle$  is given by the following integral
\begin{align}
  \langle \Phi_{\Delta_1}^{m_1}(\vec{w}_1)  \Phi_{\Delta_2}^{m_2} (\vec{w}_2) \phi_{\Delta_3}(\vec{w}_3)\rangle=i \lambda \int d^4 X\, \Phi_{\Delta_1}^{m_1(-)}\left(X ; \vec{w}_1\right)  \Phi_{\Delta_2}^{m_2(+)}\left(X ; \vec{w}_2\right)   \phi_{\Delta_3}^{(+)}\left(X ; \vec{w}_3 \right).
\end{align}
After the integration over $X$ it becomes
\begin{align}
  i \lambda (2\pi)^4 \displaystyle{ \int \left(\prod_{i=1}^{2}\frac{d y_i  d^2 z_i}{y_i^3} \left(\frac{y_i}{y_i^2+|\vec{z}_i-\vec{w}_i|^2}\right)^{\Delta_i}\right) \int \frac{d \omega_3}{\omega_3} \omega_3^{\Delta_3} \delta^4(-m_1 \hat{p}_1+ m_2\hat{p}_2+\omega_3\hat{q}_3)}.
\end{align}
The momentum conservating delta function can be evaluated as 
\begin{align}
\label{eq:delta}
\delta^4(-m_1 \hat{p}_1+ m_2\hat{p}_2+\omega_3\hat{q}_3)= \frac{y_1^4 (m_1^2 y_2^2+m_2^2(\vec{z}_2-\vec{w}_3)^2)}{m_1^3 m_2^2 (y_2^2 +(\vec{z}_2-\vec{w}_3)^2)^2} \delta(y_1-y_1^*) \delta^2(\vec{z}_1-\vec{z}_1^*)\delta(\omega_3-\omega_3^*),
\end{align}
where the factor is the Jacobian coming from solving the momentum conservation using hyperbolic coordinates. The fixed integration variables are as following
\begin{align}
y_1^* &= \frac{m_1 m_2 y_2 (y_2^2 +(\vec{z}_2-\vec{w}_3)^2)}{m_1^2 y_2^2+m_2^2(\vec{z}_2-\vec{w}_3)^2}\nonumber\\
\vec{z}_1^* &= \frac{m_2^2 (y_2^2 +(\vec{z}_2-\vec{w}_3)^2) \vec{z}_2 + y_2^2 (m_1^2 - m_2^2)\vec{w}_3}{m_1^2 y_2^2+m_2^2(\vec{z}_2-\vec{w}_3)^2}\nonumber\\
\omega_3^* &= \frac{ y_2 (m_1^2 - m_2^2)}{2 m_2 (y_2^2 +(\vec{z}_2-\vec{w}_3)^2)}.
\end{align}
Since $y\ge 0,\omega \ge 0$, the kinematics constrains the masses to be $m_1\ge m_2$. 

This integral can be simplified by the conformal symmetry~\eqref{eq:sl2c}. This conformal symmetry fixes the $\vec{w}$ dependence of the integral to be 
\begin{align}
  \langle \Phi_{\Delta_1}^{m_1}  \Phi_{\Delta_2}^{m_2}  \phi_{\Delta_3}\rangle \propto \frac{1}{|\vec{w}_{12}|^{\Delta_1+\Delta_2-\Delta_3}|\vec{w}_{13}|^{\Delta_1+\Delta_3-\Delta_2}|\vec{w}_{23}|^{\Delta_2+\Delta_3-\Delta_1}}.
\end{align}
The exact factor of this proportion can be computed by using the famous  linear fractional transformation to fix the three points on the celestial sphere 
\begin{align}
\vec{w}_1\to \infty,\, \vec{w}_2\to \vec{1}=(1,0),\, \vec{w}_3\to 0.
\end{align}
Plugging these fixed points into the integral, we  obtain the factor of proportion~\footnote{The term $1/(\infty^2)^{\Delta_1}$ coming from $\vec{w}_1\to\infty$ is cancelled by the Jacobian of this  linear fractional transformation.} as following
\begin{align}
    i \lambda (2\pi)^4 \displaystyle{ \int \left(\prod_{i=1}^{2}\frac{d y_i  d^2 z_i}{y_i^3}\right) \frac{y_1^{\Delta_1} y_2^{\Delta_2}}{(y_2^2+|\vec{z}_2-\vec{1}|^2)^{\Delta_2}} \int \frac{d \omega_3}{\omega_3}\, \omega_3^{\Delta_3}\, \frac{y_1^4 (m_1^2 y_2^2+m_2^2\vec{z}_2^2)}{m_1^3 m_2^2 (y_2^2 +\vec{z}_2^2)^2}} \nonumber\\
  \times  \delta(y_1-\frac{m_1 m_2 y_2 (y_2^2 +\vec{z}_2^2)}{m_1^2 y_2^2+m_2^2\vec{z}_2^2}) \delta^2(\vec{z}_1-\frac{m_2^2 (y_2^2 +\vec{z}_2^2) \vec{z}_2 }{m_1^2 y_2^2+m_2^2\vec{z}_2^2}) \delta(\omega_3-\frac{ y_2 (m_1^2 - m_2^2)}{2 m_2 (y_2^2 +\vec{z}_2^2)}).
\end{align}
Eliminating the delta functions we are left with a three-fold integral in a single hyperbolic space 
\begin{align}
  \label{eq:fac0}
  \frac{i \lambda (2\pi)^4 (m_1^2-m_2^2)^{\Delta_3-1}}{2^{\Delta_3-1} m_1^{2-\Delta_1} m_2^{\Delta_1+\Delta_3}} \displaystyle{ \int d y_2  d^2 z_2   \frac{y_2^{\Delta_1+\Delta_2+\Delta_3-3}}{(y_2^2+|\vec{z}_2-\vec{1}|^2)^{\Delta_2}\, (y_2^2+\vec{z}_2^2)^{\Delta_3-\Delta_1}\, (\vec{z}_2^2 + \frac{m_1^2}{m_2^2}y_2^2)^{\Delta_1}}} .
\end{align}

This integral can be computed using the Feynman parametrization. The denominator under the Feynman parametrization becomes 
\begin{align}
  \label{eq:fac1}
  &\frac{1}{(y_2^2+|\vec{z}_2-\vec{1}|^2)^{\Delta_2}\, (y_2^2+\vec{z}_2^2)^{\Delta_3-\Delta_1}\, (\vec{z}_2^2 + \frac{m_1^2}{m_2^2}y_2^2)^{\Delta_1}} = \frac{ \Gamma(\Delta_2+\Delta_3)}{ \Gamma(\Delta_2)\Gamma(\Delta_3-\Delta_1)\Gamma(\Delta_1)} \nonumber\\
  &\times \int_{0}^{1} dt_1 \int_{0}^{1-t_1} dt_2 \frac{t_1^{\Delta_2-1} t_2^{\Delta_3-\Delta_1-1} (1-t_1-t_2)^{\Delta_1-1}}{\big[(\vec{z}_2-t_1 \vec{1})^2 +t_1 (1-t_1) + y_2^2 \big(t_1+t_2+(1-t_1-t_2)\frac{m_1^2}{m_2^2}\big)\big]^{\Delta_2+\Delta_3}} .
\end{align}
Then the integration over $\vec{z}_2$ is the standard Euclidean integral encountered in dimensional regularization 
\begin{align}
  \label{eq:fac2}
\int d^2z_2 &\frac{1}{\big[(\vec{z}_2-t_1 \vec{1})^2 +t_1 (1-t_1) + y_2^2 \big(t_1+t_2+(1-t_1-t_2)\frac{m_1^2}{m_2^2}\big)\big]^{\Delta_2+\Delta_3}}  \nonumber\\
&=\frac{\pi \Gamma(\Delta_2+\Delta_3-1)}{\Gamma(\Delta_2+\Delta_3)}  \frac{1}{\big[t_1 (1-t_1) + y_2^2 \big(t_1+t_2+(1-t_1-t_2)\frac{m_1^2}{m_2^2}\big)\big]^{\Delta_2+\Delta_3-1}}.
\end{align}
The integration over $y_2$ can be computed using the Euler beta function 
\begin{align}
  \label{eq:fac3}
  \int_0^\infty dy_2 &\frac{y_2^{\Delta_1+\Delta_2+\Delta_3-3}}{\big[t_1 (1-t_1) + y_2^2 \big(t_1+t_2+(1-t_1-t_2)\frac{m_1^2}{m_2^2}\big)\big]^{\Delta_2+\Delta_3-1}}  \nonumber\\
  &=\frac{\Gamma(\frac{\Delta_1+\Delta_2+\Delta_3}{2}-1) \Gamma(\frac{\Delta_2+\Delta_3-\Delta_1}{2})}{2 \Gamma(\Delta_2+\Delta_3-1)}  \frac{\big[t_1 (1-t_1)\big]^{\frac{\Delta_1-\Delta_2-\Delta_3}{2}}}{\big[t_1 +t_2 +(1-t_1-t_2)\frac{m_1^2}{m_2^2}\big]^{\frac{\Delta_1+\Delta_2+\Delta_3}{2}-1}}.
\end{align}
Finally only the integration of Feynman parameters is left 
\begin{align}
  \label{eq:feynpara}
  \int_{0}^{1} dt_1 \int_{0}^{1-t_1} dt_2 \frac{t_1^{\frac{\Delta_1+\Delta_2-\Delta_3}{2}-1} t_2^{\Delta_3-\Delta_1-1} (1-t_1-t_2)^{\Delta_1-1} (1-t_1)^{\frac{\Delta_1-\Delta_2-\Delta_3}{2}}}{\big[t_1 +t_2 +(1-t_1-t_2)\frac{m_1^2}{m_2^2}\big]^{\frac{\Delta_1+\Delta_2+\Delta_3}{2}-1}}.
\end{align}
Using the famous Mellin-Barnes method, this integral can be computed into an infinite series of the mass ratio $m_2/m_1$ as following 
\begin{align}
  \label{eq:facMB}
  &\frac{\Gamma(\Delta_1) \Gamma(\Delta_3-\Delta_1) }{\Gamma(\frac{\Delta_1+\Delta_2+\Delta_3}{2}-1)\Gamma(\frac{\Delta_1-\Delta_2-\Delta_3}{2}+1)\Gamma(\Delta_3)} \sum_{n=0}^{\infty}\, \Bigg\{ \frac{(-1)^n}{n!} \left(\frac{m_2}{m_1}\right)^{\Delta_1+\Delta_2+\Delta_3-2 +2 n}  \nonumber\\
&\times \Gamma(1-\Delta_2-n) \Gamma(\frac{\Delta_1+\Delta_2-\Delta_3}{2}+n) \Gamma(\frac{\Delta_1+\Delta_2+\Delta_3}{2}-1+n)  
\nonumber\\
&\times  { }_2 F_1\left(\begin{array}{c}
  \Delta_3-\Delta_1, \frac{\Delta_1+\Delta_2+\Delta_3}{2}-1+n \\
  \Delta_3
\end{array}; 1- \frac{m_2^2}{m_1^2}\right) 
+ (\Delta_2 \rightarrow \tilde{\Delta}_2) \Bigg\} ,
\end{align}
To keep focused on the main physics, we put the computation of this series into the appendix~\footnote{Note that it is hard to carry out the summation of this series into  a concise and clean expression. }. 
We see that a special gamma function $\Gamma(1-\Delta_2-n)$ appears, which contributes a pole  $\Delta_2\to 1-n$. Similarly for $\tilde{\Delta}_2$. This is the key result of our paper and these poles are crucial for the massless limit of this 3pt amplitude, which is discussed in the next section.

\section{\label{sec:softlimit}Massless limit and operator dimensions}

For this 3pt massive-massive-massless amplitude, the massless limit $m_2\to 0$ makes sense physically. The kinematics only requires the masses to satisfy $m_1 \ge m_2$. It is legitimate to imagine situations where the outgoing particle  becomes lighter and lighter. As long as $m_2 < m_1$, no abrupt change of physical process happens. So it is safe to imagine that when $m_2\to 0$, the resulting amplitude becomes the massive-massless-massless amplitude. 

The massless limit of the amplitude should be consistent with the massless limit of the conformal primary waves~\eqref{eq:basislimit}, where $\Phi_{\Delta_2}^{m_2}$ has two limiting massless fields  $\phi_{\Delta_2}$ and $\reallywidetilde{\phi_{\tilde{\Delta}_2}}$. So the  massless limit of the 3pt amplitude should contain two parts, which are as follows after combining  with the phase factor in~\eqref{eq:basislimit}
\begin{align}
	\label{eq:match}
	&\left(\frac{m_2}{2}\right)^{\Delta_2} \frac{\Gamma\left(\Delta_2\right)} {\pi\Gamma(\Delta_2-1)}  \langle \Phi_{\Delta_1}^{m_1} \Phi_{\Delta_2}^{m_2} \phi_{\Delta_3}\rangle \overunderset{?}{m_2\to 0}{\Longrightarrow} \langle \Phi_{\Delta_1}^{m_1} \phi_{\Delta_2} \phi_{\Delta_3}\rangle, \nonumber\\
	&\left(\frac{m_2}{2}\right)^{\tilde{\Delta}_2} \frac{\Gamma\left(\tilde{\Delta}_2\right)} {\pi\Gamma(\tilde{\Delta}_2-1)}  \langle \Phi_{\Delta_1}^{m_1} \Phi_{\Delta_2}^{m_2} \phi_{\Delta_3}\rangle \overunderset{?}{m_2\to 0}{\Longrightarrow} \langle \Phi_{\Delta_1}^{m_1} \reallywidetilde{\phi_{\tilde{\Delta}_2}} \phi_{\Delta_3}\rangle.
\end{align}
where the question mark means this equality has to be proved. Since all amplitudes have the same functional dependence on $\vec{w}$, we only need to check the factor of proportion.

\subsection{\label{sec:leading}Leading term of $m_2\to 0$ and conformally soft  $\Delta_2\to 1$}

We firsly check the leading term of the massless limit. The subleading orders  are different and details will be  discussed in section~\ref{sec:subleading}.

\textbf{[I]:} For the case of  $\phi_{\Delta_2}$,  combining the factors in ~\eqref{eq:fac0}-\eqref{eq:fac3} and~\eqref{eq:facMB}, the left hand side of~\eqref{eq:match} under $m_2\to 0$ is 
\begin{align}
  i\lambda (2\pi)^4\frac{m_1^{\Delta_3-\Delta_2-2}}{2^{\Delta_2+\Delta_3} m_2^{2-2\Delta_2}}\frac{\Gamma(1-\Delta_2) }{\Gamma(\Delta_2-1) } \frac{\Gamma(\frac{\Delta_1+\Delta_2+\Delta_3}{2}-1) \Gamma(\frac{\Delta_1+\Delta_2-\Delta_3}{2})\Gamma(\frac{\Delta_2+\Delta_3-\Delta_1}{2})}{\Gamma(\Delta_1) \Gamma(1+ \frac{\Delta_3-\Delta_1-\Delta_2}{2})}. 
\end{align}
This is a well-defined finite limit only if $\Delta_2\to 1$. So the massless limit $m_2\to 0$ requires the conformal soft limit $\Delta_2=1$, if decreasing $m_2$ is a well-behaved physical process.
We see that the numerator $\Gamma(1-\Delta_2)$ coming from the massive amplitude is crucial in the conformal soft limit. It gives the leading contribution in the combined limit. The denominator $\Gamma(\Delta_2-1)$ coming from the primary wave function can be viewed as a 'normalization', which balances the pole of the numerator and thus making the final amplitude well-defined. 

Now with both $m_2\to 0$ and $\Delta_2=1$, the factors on both sides of~\eqref{eq:match} take the same value~\footnote{There is a total minus sign in the left hand side of~\eqref{eq:match}, which is irrelevant for the physics discussion in our paper. }
\begin{align}
  i\lambda (2\pi)^4\frac{m_1^{\Delta_3-3}}{2^{\Delta_3+1} }  \frac{\Gamma(\frac{\Delta_1+\Delta_3-1}{2}) \Gamma(\frac{\Delta_1+1-\Delta_3}{2})}{\Gamma(\Delta_1)}.
\end{align}
This concludes the finding of this paper
\begin{align}
  \label{eq:core}
  \left(\frac{m_2}{2}\right)^{\Delta_2} \frac{\Gamma\left(\Delta_2\right)} {\pi\Gamma(\Delta_2-1)}  \langle \Phi_{\Delta_1}^{m_1} \Phi_{\Delta_2}^{m_2} \phi_{\Delta_3}\rangle \xRightarrow[m_2\to 0]{\Delta_2\to 1} \langle \Phi_{\Delta_1}^{m_1} \phi_{\Delta_2=1} \phi_{\Delta_3}\rangle.
\end{align}

\textbf{[II]:} For the case of  $\reallywidetilde{\phi_{\tilde{\Delta}_2}}$, the left hand side of~\eqref{eq:match} under $m_2\to 0$ is 
\begin{align}
	i\lambda (2\pi)^4\frac{m_1^{\Delta_3-\tilde{\Delta}_2-2}}{2^{\tilde{\Delta}_2+\Delta_3} m_2^{2-2\tilde{\Delta}_2}}\frac{\Gamma(1-\tilde{\Delta}_2) }{\Gamma(\tilde{\Delta}_2-1) } \frac{\Gamma(\tilde{\Delta}_2) \Gamma(\frac{\Delta_1+\tilde{\Delta}_2+\Delta_3}{2}-1) \Gamma(\frac{\Delta_1-\tilde{\Delta}_2-\Delta_3}{2}+1)}{\Gamma(\Delta_1) \Gamma(\Delta_2) }. 
\end{align}
Similarly, this is a well-defined finite limit of $m_2\to 0$ only if $\tilde{\Delta}_2\to 1$. Now both sides of~\eqref{eq:match} take the same value
\begin{align}
	i\lambda (2\pi)^4\frac{m_1^{\Delta_3-3}}{2^{\Delta_3+1} }  \frac{\Gamma(\frac{\Delta_1+\Delta_3-1}{2}) \Gamma(\frac{\Delta_1+1-\Delta_3}{2})}{\Gamma(\Delta_1)}.
\end{align}
So we also get 
\begin{align}
	\label{eq:coreShadow}
	\left(\frac{m_2}{2}\right)^{\tilde{\Delta}_2} \frac{\Gamma\left(\tilde{\Delta}_2\right)} {\pi\Gamma(\tilde{\Delta}_2-1)}  \langle \Phi_{\Delta_1}^{m_1} \Phi_{\Delta_2}^{m_2} \phi_{\Delta_3}\rangle \xRightarrow[m_2\to 0]{\tilde{\Delta}_2\to 1} \langle \Phi_{\Delta_1}^{m_1} \reallywidetilde{\phi_{\tilde{\Delta}_2=1}} \phi_{\Delta_3}\rangle.
\end{align}
Note that at the leading order of the massless limit, we obtained $\Delta_2=\tilde{\Delta}_2=1$ and the two amplitudes are the same. This is consistent because the unshadowed and the shadowed field are the same in the $\Delta_2 =1$ limit~\cite{Pasterski:2017kqt}.

\subsection{\label{sec:subleading}Subleading term $m_2^{2n}$ and analytically continued $\Delta_2=1-n$}

For convenience, we give the complete result of the 3pt amplitude here by combining the factors of ~\eqref{eq:fac0}-\eqref{eq:fac3} with~\eqref{eq:facMB}
\begin{align}
  \label{eq:totAmp}
  &\langle \Phi_{\Delta_1}^{m_1} \Phi_{\Delta_2}^{m_2} \phi_{\Delta_3}\rangle  =  i\lambda (2\pi)^4 \frac{(m_1^2-m_2^2)^{\Delta_3-1}}{2^{\Delta_3} m_1^{2-\Delta_1} m_2^{\Delta_1+\Delta_3}}   \frac{\pi \Gamma(\frac{\Delta_2+\Delta_3-\Delta_1}{2})}{\Gamma(\Delta_2)\Gamma(\Delta_3) \Gamma(\frac{\Delta_1-\Delta_2-\Delta_3}{2}+1)} \nonumber\\
  &\times \sum_{n=0}^{\infty}\, \Bigg\{ \frac{(-1)^n}{n!} \left(\frac{m_2}{m_1}\right)^{\Delta_1+\Delta_2+\Delta_3-2 +2 n}  \Gamma(1-\Delta_2-n) \Gamma(\frac{\Delta_1+\Delta_2+\Delta_3}{2}-1+n)  \nonumber\\
  &\times  \Gamma(\frac{\Delta_1+\Delta_2-\Delta_3}{2}+n)
  { }_2 F_1\left(\begin{array}{c}
  \Delta_3-\Delta_1, \frac{\Delta_1+\Delta_2+\Delta_3}{2}-1+n \\
  \Delta_3
\end{array}; 1- \frac{m_2^2}{m_1^2}\right) 
+ (\Delta_2 \rightarrow \tilde{\Delta}_2) \Bigg\} \nonumber \\
&\propto  \ldots \frac{1}{\Gamma(\Delta_2)}\sum_{n=0}^{\infty} \Bigg\{ m_2^{\Delta_2-(2-2 n)}  \Gamma(1-n-\Delta_2) \ldots + m_2^{-\Delta_2+2 n}  \Gamma(\Delta_2 -1-n)\ldots \Bigg\},
\end{align}
where the sum is symmetric under the switch $\Delta_2 \leftrightarrow \tilde{\Delta}_2$ and we emphasize  the key quantities in the last line. Note that the total amplitude is not symmetric under the switch $\Delta_2 \leftrightarrow \tilde{\Delta}_2$, because it is an amplitude of $ \Phi_{\Delta_2}^{m_2}$, not of  $ \Phi_{\Delta_2}^{m_2} +    \Phi_{\tilde{\Delta}_2}^{m_2}$. Then it is obvious that the poles of the two parts of the sum can not contribute in a 'symmetric' way to the total amplitude, although they are symmetric inside the sum. The $1/\Gamma(\Delta_2)$ plays the key role in this distinction between the  poles of the two parts. 

Now we continue to analyze the subleading orders of the massless limit. In~\eqref{eq:completeIntegral}, there is an infinite number of subleading terms $m_2^{2n},$ $n>0$, associated with functions $\Gamma(1-\Delta_2-n)$ and $\Gamma(1-\tilde{\Delta}_2-n)$. They can also contribute poles when the scaling dimensions are analytically continued  to $\Delta_2= 1-n$ or $\tilde{\Delta}_2= 1-n$. Now at subleading orders we can distinguish between the two phase factors of  $\phi_{\Delta_2}$ and $\reallywidetilde{\phi_{\tilde{\Delta}_2}}$ in~\eqref{eq:basislimit}. The poles coming from $\Gamma(1-\Delta_2-n)$ or $\Gamma(1-\tilde{\Delta}_2-n)$ must be cancelled by the denominator of the phase factors $\Gamma(\Delta_2-1)$ or $\Gamma(\tilde{\Delta}_2-1)$ respectively. 

\textbf{[I]:} Firstly let's consider the case of  $\phi_{\Delta_2}$ that we used in~\eqref{eq:match}. Let's pick the n-th term from the series~\eqref{eq:completeIntegral} by analytically continued  dimension $\Delta_2= 1-n$. The first part of the sum (terms of dimension $\Delta_2$) captures the corresponding subleading term of $m_2$, but the second part of the sum (terms of dimension $\tilde{\Delta}_2$) has an extra factor $m_2^{2n}$ that is much more subleading. This is shown in the following
\begin{align}
  \label{eq:case1-sk}
	\langle \Phi_{\Delta_1}^{m_1} \Phi_{\Delta_2}^{m_2} \phi_{\Delta_3}\rangle  &\propto  \frac{1}{\Gamma(1-n)} \Bigg\{ m_2^{-1+n} \ldots +   m_2^{-1+3 n}\ldots\Bigg\}  \nonumber\\
	&\propto    \phi_{\Delta_2=1-n}  + m_2^{4 n}  \frac{1}{\Gamma(1-n) } \tilde{\phi}_{\tilde{\Delta}_2=1+n}\,(\textrm{using ~\eqref{eq:basislimit}}) \nonumber\\
  &	\propto    \phi_{\Delta_2=1-n} . 
\end{align}
 So the contribution of the second part (of dimension $\tilde{\Delta}_2$) vanishes at the specific subleading order of the massless limit. In detail, the first part of the sum contributes to the left hand side of~\eqref{eq:match} as
\begin{align}
  i\lambda (2\pi)^4& \frac{m_1^{\Delta_3-\Delta_2-2-2n}}{2^{\Delta_2+\Delta_3} m_2^{2-2\Delta_2-2n}}\frac{\Gamma(1-\Delta_2-n) }{\Gamma(\Delta_2-1) } \frac{(-1)^n}{n!} \nonumber\\
  &\times \frac{\Gamma(\frac{\Delta_1+\Delta_2+\Delta_3}{2}-1+n) \Gamma(\frac{\Delta_2+\Delta_3-\Delta_1}{2}) \Gamma(\frac{\Delta_1+\Delta_2-\Delta_3}{2}+n)\Gamma(\frac{\Delta_1-\Delta_2-\Delta_3}{2}+1-n)}{\Gamma(\Delta_1) \Gamma(\frac{\Delta_1-\Delta_2-\Delta_3}{2}+1)\Gamma(1-n + \frac{\Delta_3-\Delta_1-\Delta_2}{2})}. 
\end{align}
When the scaling dimension is analytically continued to $\Delta_2=1-n$, this contribution is finite and well-behaved as 
\begin{align}
   i\lambda (2\pi)^4 \frac{m_1^{\Delta_3+\Delta_2-4}}{2^{\Delta_2+\Delta_3}} \frac{\Gamma(\frac{\Delta_1+\Delta_2-\Delta_3}{2}) \Gamma(\frac{\Delta_1-\Delta_2+\Delta_3}{2})}{\Gamma(\Delta_1) } .
\end{align}
This means that the subleading term $m_2^{2n}$ combined with analytically continued $\Delta_2=1-n$ leads to the 3pt amplitude~\footnote{For subleading terms, there is a subtlety here. The analytic continuation $\Delta_2=1-n$ has to be taken before the massless limit $m_2\to 0$, because they have preceding terms $1, m_2^2,\ldots, m_2^{2n-2}$ that are more singular in the massless limit. This subtlety is absent for the leading term. }
\begin{align}
  \label{eq:subordermatch}
  \left(\frac{m_2}{2}\right)^{\Delta_2} \frac{\Gamma\left(\Delta_2\right)} {\pi\Gamma(\Delta_2-1)}  \langle \Phi_{\Delta_1}^{m_1} \Phi_{\Delta_2}^{m_2} \phi_{\Delta_3}\rangle \xRightarrow[m_2\to 0]{\Delta_2\to 1-n}   \langle \Phi_{\Delta_1}^{m_1} \phi_{\Delta_2=1-n} \phi_{\Delta_3}\rangle. 
\end{align}
The nontrivial massless limit at these subleading orders does give the exact 3pt amplitude $\langle \Phi_{\Delta_1}^{m_1} \phi_{\Delta_2} \phi_{\Delta_3}\rangle$. We see that the analytic continuation on scaling dimensions is a powerful tool in selecting subleading contributions in the massless limit of massive amplitude. It happens that these analytically continued dimensions belong to the range of generalized conformal primary operators $\Delta \in 2-\mathbb{Z}_{\geqslant 0}$ of massless bosons~\cite{Mitra:2024ugt}. 

\textbf{[II]:} Secondly let's consider the case of $\reallywidetilde{\phi_{\tilde{\Delta}_2}}$ to see if the following equality works
\begin{align}
  \label{eq:shadow}
  \left(\frac{m_2}{2}\right)^{\tilde{\Delta}_2} \frac{\Gamma\left(\tilde{\Delta}_2\right)} {\pi\Gamma(\tilde{\Delta}_2-1)}  \langle \Phi_{\Delta_1}^{m_1} \Phi_{\Delta_2}^{m_2} \phi_{\Delta_3}\rangle \overunderset{?}{m_2\to 0}{\Longrightarrow} \langle \Phi_{\Delta_1}^{m_1} \reallywidetilde{\phi_{\tilde{\Delta}_2}} \phi_{\Delta_3}\rangle.
\end{align}
Now pick the n-th term from the series~\eqref{eq:completeIntegral} by analytically continued  dimension $\tilde{\Delta}_2= 1-n$. The first part of the sum (terms of dimension $\Delta_2$) has an extra factor $m_2^{2n}$ that is much more subleading, and only the second part of the sum (terms of dimension $\tilde{\Delta}_2$) captures the corresponding subleading term. This is shown in the following
\begin{align}
	\langle \Phi_{\Delta_1}^{m_1} \Phi_{\Delta_2}^{m_2} \phi_{\Delta_3}\rangle  &\propto  \frac{1}{\Gamma(1+n)} \Bigg\{  m_2^{-1+3n} \ldots + m_2^{-1+n}\ldots \Bigg\} \nonumber\\
	&\propto  m_2^{4 n}  \phi_{\Delta_2=1+n}  + \frac{1}{\Gamma(1+ n)  }\tilde{\phi}_{\tilde{\Delta}_2=1-n} \,(\textrm{using ~\eqref{eq:basislimit}}) \nonumber\\
	&\propto  \frac{1}{\Gamma(1+ n)  }  \tilde{\phi}_{\tilde{\Delta}_2=1-n} . 
\end{align}
Note that in the second step we should use the factor associated with the shadowed field in ~\eqref{eq:basislimit}. 
So the contribution of the first part (of dimension $\Delta_2$) vanishes at the specific subleading order of the massless limit.
We can also see that the subleading terms are not exactly the massless primaries, because of the overall factor. Only for the subleading order $n=2$ we have the massless primary because $\Gamma(2)=1$. 

Combining it with all the factors,  the second part of the sum contributes to the left hand side of~\eqref{eq:shadow} as
\begin{align}
 & \frac{(-1)^{n+1} i\lambda (2\pi)^4  m_1^{\Delta_3+\tilde{\Delta}_2-4}}{(n-1)!  2^{\tilde{\Delta}_2+\Delta_3}} \frac{\Gamma(\frac{\Delta_1+\Delta_2-\Delta_3}{2}) \Gamma(\frac{\Delta_2+\Delta_3-\Delta_1}{2})\Gamma(-1+\frac{\Delta_1+\Delta_2+\Delta_3}{2})}{\Gamma(\Delta_2) \Gamma(\Delta_1) \Gamma(1+\frac{\Delta_3-\Delta_1-\Delta_2}{2})}.
\end{align}
Compared with the 3pt shadowed amplitude of~\eqref{eq:3ptshadow},  we obtain the massless limit 
\begin{align}
  \label{eq:shadowlimit}
  \left(\frac{m_2}{2}\right)^{\tilde{\Delta}_2} \frac{\Gamma\left(\tilde{\Delta}_2\right)} {\pi\Gamma(\tilde{\Delta}_2-1)}  \langle \Phi_{\Delta_1}^{m_1} \Phi_{\Delta_2}^{m_2} \phi_{\Delta_3}\rangle \overunderset{\overset{\tilde{\Delta}_2\to 1-n}{\Delta_2\to 1+n}}{m_2\to 0}{\Longrightarrow} \frac{(-1)^{n+1} }{(n-1)!  \Gamma(\Delta_2=1+n)}  \langle \Phi_{\Delta_1}^{m_1} \reallywidetilde{\phi_{\tilde{\Delta}_2=1-n}} \phi_{\Delta_3}\rangle.
\end{align}
There is a mismatch by the strange factor $1/(n-1)! \Gamma(\Delta_2=1+n)$ at the n-th subleading order. Only the first subleading order $n=1$ gives an exact match, whose operator dimension is $\Delta_2 =2$. 
This is easy to understand: the total amplitude~\eqref{eq:totAmp} is not symmetric under the switch $\Delta_2 \leftrightarrow \tilde{\Delta}_2$, because of the overall prefactor $1/\Gamma(\Delta_2)$. 

All other subleading orders with dimension $\Delta_2 \ge 3$ do not give an exact match, and they happen to be out of the range of generalized conformal primary operators $\Delta \in 2-\mathbb{Z}_{\geqslant 0}$ of massless bosons~\cite{Mitra:2024ugt}. So the subleading orders of the massless limit associated with  $\reallywidetilde{\phi_{\tilde{\Delta}_2}}$ manifests the constraint of generalized conformal primary operators $\Delta \in 2-\mathbb{Z}_{\geqslant 0}$. Only for the generalized conformal primary operator it can reduce to the exact 3pt amplitude $\langle \Phi_{\Delta_1}^{m_1} \reallywidetilde{\phi_{\tilde{\Delta}_2}} \phi_{\Delta_3}\rangle$.

\subsection{Vanish of the 3pt celestial amplitude $\langle \phi_{\Delta_1} \phi_{\Delta_2} \phi_{\Delta_3}\rangle$}

We further check the massless limit and conformal soft limit. Take the 3pt amplitude 
\begin{align}
  \langle \Phi_{\Delta_1}^{m_1} \phi_{\Delta_2} \phi_{\Delta_3}\rangle \propto \left(m_1\right)^{\Delta_2+\Delta_3-4} \frac{\Gamma(\frac{\Delta_1+\Delta_2-\Delta_3}{2}) \Gamma(\frac{\Delta_1-\Delta_2+\Delta_3}{2})}{\Gamma(\Delta_1)}. 
\end{align}
This amplitude does not have a function like $\Gamma(1-\Delta_1)$, so there is no pole to sustain the amplitude in combined limit. Take the massless limit $m_1\to 0$ 
\begin{align}
  \langle \phi_{\Delta_1} \phi_{\Delta_2} \phi_{\Delta_3}\rangle \propto  \left(m_1\right)^{\Delta_1} \frac{\Gamma\left(\Delta_1\right)} {\pi\Gamma(\Delta_1-1)} \langle \Phi_{\Delta_1}^{m_1} \phi_{\Delta_2} \phi_{\Delta_3}\rangle.
\end{align}
In the conformal soft limit $\Delta_1\to 1$, it is either zero or infinite
\begin{align}
  \langle \phi_{\Delta_1} \phi_{\Delta_2} \phi_{\Delta_3}\rangle \propto  i\lambda_1 \left(m_1\right)^{-1+i\lambda_1+i\lambda_2+i\lambda_3},
\end{align}
depending on which one of $m_1$ and $\lambda_1$ goes to zero faster. 
The zero or infinity means that it does not have a limiting amplitudes of three massless scalars. This might be another explanation of the fact that 3pt celestial massless amplitudes vanish due to 4d kinematics.

\section{Conclusion}

In this work, we studied the massless limit of the 3pt celestial amplitude of two massive states $\langle \Phi_{\Delta_1}^{m_1} \Phi_{\Delta_2}^{m_2} \phi_{\Delta_3}\rangle$. In the massless limit $m_2\to 0$, it reduces to the 3pt celestial amplitude of one massive state $\langle \Phi_{\Delta_1}^{m_1} \phi_{\Delta_2=1} \phi_{\Delta_3}\rangle$, provided with the conformal soft limit $\Delta_2\to 1$. The pole $1/(1-\Delta_2)$ coming from $\Gamma(1-\Delta_2)$ of the massless limit are crucial for the physics here.  This can be compared with the conformal soft limit of celestial gluons, where the soft energy $\omega\to 0$ gives the leading contribution and a pole of $1/(\Delta- 1)$ arises. For celestial gluons, higher-point amplitudes are used, because the 3pt amplitude vanishes due to 4d kinematics and the amplitudes reduce to lower-point amplitudes during the conformal soft limit. Here, the presence of massive states preserves the 3pt amplitude and it does not reduce to lower-point amplitude. 

This connection between the $m_2\to 0$ and the $\Delta_2\to 1$ is unexpected and interesting, because naively one expects that the resulting amplitude under the $m_2\to 0$ limit can have arbitrary dimension $\Delta_2$. But the results from this celestial amplitude show that it is only consistent for the very special dimension $\Delta_2 =1$, which is related with the conformal soft modes of massless primaries. To understand the detailed mechanism of this connection, a new and independent analysis is necessary that is different from the celestial amplitude approach, for example, an analysis from the asymptotic symmetry.  This is one of the open questions. 

We also find that the analytic continuation of scaling dimensions is a powerful tool in selecting subleading orders of the massless limit of massive amplitudes. These scaling dimensions $\Delta_2=1-n$ and $\Delta_2=2$ fall into the range of  generalized conformal primary operators $\Delta \in 2-\mathbb{Z}_{\geqslant 0}$ of massless bosons~\cite{Mitra:2024ugt}, which are also the scaling dimensions of the $\textrm{w}_{1+\infty}$ algebra~\cite{Strominger:2021mtt}. These generalized conformal primaries are obtained from analytic constraints on the massless primaries and the $\textrm{w}_{1+\infty}$ algebra is from the current algebra of asymptotic symmetry. It is interesting that these generalized conformal primaries are recovered from the massless limits of massive amplitude. Can we dig more from this connection? This is another open question.

It would be interesting and meaningful to know what would happen when all three particles are massive $\langle \Phi_{\Delta_1}^{m_1} \Phi_{\Delta_2}^{m_2} \Phi_{\Delta_3}^{m_3}\rangle$. When taking the massless limit, for example $m_3\to 0$, would it reduce to the 3pt amplitude $\langle \Phi_{\Delta_1}^{m_1} \Phi_{\Delta_2}^{m_2} \phi_{\Delta_3}\rangle$ together with $\Delta_3\to 1$? However, this 3pt celestial amplitude of three massive scalars are hard to compute. After eliminating the momentum conservating delta function, a five-fold integral is left. In the near-extremal limit $m_1=2(1+\epsilon)m$, $m_2=m_3=m$, $\epsilon\to 0$, the five-fold integral reduces  to a three-fold integral in a single hyperbolic space, which gives the tree-level 3pt Witten diagram at the leading order of $\sqrt{\epsilon}$~\cite{Pasterski:2016qvg}. For general mass configurations, the five-fold integral of hyperbolic coordinates is very complicated. Furthermore, two of the five integration variables are coupled and have nontrivial end points in their integration region depending on the masses. Currently this five-fold integral of hyperbolic coordinates remains to be an open problem. 

\backmatter
\bmhead{Acknowledgements}
 Wei Fan is supported in part by the National Natural Science Foundation of China under Grant No.\ 12105121.

\begin{appendices}
\section{The integral of Feynman parameters for general masses}
%\label{sec:appendixA}
In this appendix, we compute the integral of Feynman parameters~\eqref{eq:feynpara}. 
Firstly we do a change of variable 
\begin{align}
t_2 = (1-t_1) s,\quad s\in[0, 1],
\end{align}
to decouple the integration regions of this double-integral. This leads to 
\begin{align}
  \int_{0}^{1} dt_1 \int_{0}^{1} ds\, \frac{t_1^{\frac{\Delta_1+\Delta_2-\Delta_3}{2}-1} s^{\Delta_3-\Delta_1-1} (1-t_1)^{-\Delta_2} (1-s)^{\Delta_1- 1}}{\big[\frac{t_1}{1-t_1} +s +(1-s)\frac{m_1^2}{m_2^2}\big]^{\frac{\Delta_1+\Delta_2+\Delta_3}{2}-1}}.
\end{align}
Then the denominator can be rewritten using the famous Mellin-Barnes formula to decouple the integrand of this double-integral
\begin{align}
  \frac{1}{\big[\frac{t_1}{1-t_1} +s +(1-s)\frac{m_1^2}{m_2^2}\big]^{\frac{\Delta_1+\Delta_2+\Delta_3}{2}-1}}=&\frac{1}{\Gamma(\frac{\Delta_1+\Delta_2+\Delta_3}{2}-1)}\displaystyle{\int_{-i\infty}^{i\infty} \frac{d\alpha}{2\pi i} } \frac{(t_1/(1-t_1))^{\alpha}}{\big[s +(1-s)\frac{m_1^2}{m_2^2}\big]^{\frac{\Delta_1+\Delta_2+\Delta_3}{2}-1+\alpha}}\nonumber\\
&\times{}  \Gamma(\frac{\Delta_1+\Delta_2+\Delta_3}{2}-1+\alpha)\Gamma(-\alpha),
\end{align}
with the integration contour to be  $-1/2 < \operatorname{Re}(\alpha) <0$. 

Now the two integrals of $t_1$ and $s$ can be done independently. The $t_1$ integration gives
\begin{align}
  \int_{0}^{1} dt_1 \, \frac{t_1^{\frac{\Delta_1+\Delta_2-\Delta_3}{2}-1+\alpha} }{(1-t_1)^{-\Delta_2-\alpha} } = \frac{\Gamma(1-\Delta_2-\alpha) \Gamma(\frac{\Delta_1+\Delta_2-\Delta_3}{2}+\alpha)}{\Gamma(\frac{\Delta 1-\Delta 2-\Delta 3}{2}+1)}.
\end{align} 
And the $s$ integration gives 
\begin{align}
 \int_{0}^{1} ds\, &\frac{s^{\Delta_3-\Delta_1-1}   (1-s)^{\Delta_1- 1}}{\big[s +(1-s)\frac{m_1^2}{m_2^2}\big]^{\frac{\Delta_1+\Delta_2+\Delta_3}{2}-1+\alpha}} =\frac{\Gamma(\Delta_1) \Gamma(\Delta_3-\Delta_1)}{\Gamma(\Delta_3)} \nonumber\\
 &\times  \left(\frac{m_2}{m_1}\right)^{\Delta_1+\Delta_2+\Delta_3-2 +2\alpha}  { }_2 F_1\left(\begin{array}{c}
  \Delta_3-\Delta_1, \frac{\Delta_1+\Delta_2+\Delta_3}{2}-1+\alpha \\
  \Delta_3
\end{array}; 1- \frac{m_2^2}{m_1^2}\right).
\end{align} 
Finally we perform the Mellin-Barnes integral of $\alpha$. Because the scaling dimensions are on the principal continuous series $\Delta = 1+ i\lambda, \lambda\in\mathbb{R}$, the poles in the integrand are well separated from each other, so there is no pinched singularity to worry about. We can safely close the contour to the right hand side of the complex plane of $\alpha$, which picks the poles of $\Gamma(-\alpha)$ and  $\Gamma(1-\Delta_2-\alpha)$. The contribution of  $\Gamma(-\alpha)$ and $\Gamma(1-\Delta_2-\alpha)$ generates terms of dimension $\Delta_2$ and $\tilde{\Delta}_2 = 2 - \Delta_2$ respectively, and these terms turn out to be the same under the switch of dimension $\Delta_2 \leftrightarrow \tilde{\Delta}_2$. So the contribution of $\Gamma(-\alpha)$ and $\Gamma(1-\Delta_2-\alpha)$ are from the fields $\phi_{\Delta_2}$ and $\reallywidetilde{\phi_{\tilde{\Delta}_2}}$ respectively. 
This leads to the final result
\begin{align}
  \label{eq:completeIntegral}
  &\frac{\Gamma(\Delta_1) \Gamma(\Delta_3-\Delta_1) }{\Gamma(\frac{\Delta_1+\Delta_2+\Delta_3}{2}-1)\Gamma(\frac{\Delta_1-\Delta_2-\Delta_3}{2}+1)\Gamma(\Delta_3)} \sum_{n=0}^{\infty}\, \Bigg\{ \frac{(-1)^n}{n!} \left(\frac{m_2}{m_1}\right)^{\Delta_1+\Delta_2+\Delta_3-2 +2 n}  \nonumber\\
&\times \Gamma(1-\Delta_2-n) \Gamma(\frac{\Delta_1+\Delta_2-\Delta_3}{2}+n) \Gamma(\frac{\Delta_1+\Delta_2+\Delta_3}{2}-1+n)  
\nonumber\\
&\times  { }_2 F_1\left(\begin{array}{c}
  \Delta_3-\Delta_1, \frac{\Delta_1+\Delta_2+\Delta_3}{2}-1+n \\
  \Delta_3
\end{array}; 1- \frac{m_2^2}{m_1^2}\right) 
+ (\Delta_2 \rightarrow \tilde{\Delta}_2) \Bigg\} ,
\end{align}
where the second part of the sum is obtained from the first part by replacing $\Delta_2$ with $\tilde{\Delta}_2$. 
At the leading order of the massless limit $m_2\to 0$, only the $n=0$ term contributes and these two parts are the same under the conformal soft limit $\Delta_2 = \tilde{\Delta}_2 =1$. When the scaling dimension is analytically continued to $\Delta_2=1-n$ or $\tilde{\Delta}_2=1-n$ respectively,   subleading orders of the massless limit will be picked up, which is discussed in section~\ref{sec:subleading}.

\end{appendices}

% \section{Some title}
% Please always give a title also for appendices.

% \paragraph{Note added.} This is also a good position for notes added
% after the paper has been written.

\bibliography{massive_soft}

\end{document}